\documentclass{ifacconf}
\usepackage{tikz}
\usetikzlibrary{shapes,arrows,positioning,calc}
\usepackage{subfig}
\usepackage{graphicx}      
\usepackage{natbib}        
\usepackage{amsmath}
\usepackage{amssymb}
\usepackage{booktabs}
\begin{document}
\begin{frontmatter}

\title{Hierarchical RL-MPC for Demand Response Scheduling\thanksref{footnoteinfo}} 

\thanks[footnoteinfo]{M Bloor acknowledges funding provided by the Engineering \& Physical Sciences Research Council, United Kingdom, through grant code  EP/W524323/1. C Tsay acknowledges support from a BASF/Royal Academy of Engineering Senior Research Fellowship.}

\author[First]{Maximilian Bloor} 
\author[First]{Ehecatl Antonio Del Rio Chanona}
\author[Second]{Calvin Tsay} 
\address[First]{Imperial College London, Sargent Centre for Process Systems Engineering, South Kensington Campus London, SW7 2AZ}
\address[Second]{Imperial College London, Department of Computing, South Kensington Campus London, SW7 2AZ}

\begin{abstract}                
This paper presents a hierarchical framework for demand response optimization in air separation units (ASUs) that combines reinforcement learning (RL) with linear model predictive control (LMPC). We investigate two control architectures: a direct RL approach and a control-informed methodology where an RL agent provides setpoints to a lower-level LMPC. The proposed RL-LMPC framework demonstrates improved sample efficiency during training and better constraint satisfaction compared to direct RL control. Using an industrial ASU case study, we show that our approach successfully manages operational constraints while optimizing electricity costs under time-varying pricing. Results indicate that the RL-LMPC architecture achieves comparable economic performance to direct RL while providing better robustness and requiring fewer training samples to converge. The framework offers a practical solution for implementing flexible operation strategies in process industries, bridging the gap between data-driven methods and traditional control approaches.
\end{abstract}


\end{frontmatter}

\section{Introduction}

The growing proportion of renewable energy sources in power grids has led to greater volatility in electricity prices and supply. This creates both challenges and opportunities for energy-intensive industries. Demand response (DR) programs, which incentivize consumers to adjust their electricity usage in response to grid conditions, have emerged as a crucial tool for maintaining grid stability and efficiency. In this context, developing and operating flexible processes that can modulate their power consumption without compromising product quality or operational safety has become a key focus.  
Air separation units (ASUs), which produce high-purity oxygen, nitrogen, and argon from atmospheric air, are prime candidates for implementing DR strategies. These facilities consume significant amounts of electricity, amounting to 2.55\% of U.S. manufacturing sector electricity (\citeauthor{pattison_optimal_2016}, \citeyear{pattison_optimal_2016}), and their products can be stored cryogenically, allowing for temporal decoupling of production and demand. However, the complex dynamics and strict operational constraints of ASUs pose challenges for DR implementation. 

Optimization-based methods have been widely studied for ASU scheduling and control in DR contexts. Recent work has explored top-down approaches, e.g., using scale-bridging models with linear MPC (LMPC) and bottom-up strategies, e.g., employing economic nonlinear MPC (eNMPC). While these methods have shown promise, they face limitations in practical implementation. They rely on accurate process models, which can be challenging to develop and maintain for ASUs~\citep{tsay2019optimal}. Additionally, computational solution of large-scale optimization problems in real time can be prohibitive (\citeauthor{caspari_integration_2020}, \citeyear{caspari_integration_2020}).
Reinforcement learning (RL) offers an alternative, data-driven approach that can potentially overcome some of these limitations. RL agents can learn to approximate optimal control policies directly from process data and interactions. Once trained, RL policies generally enable quick real-time inference, avoiding the computational burden of online optimization. However, the direct application of RL to complex chemical processes such as ASUs faces challenges in terms of sample efficiency, stability, and constraint satisfaction.
In this work, we propose a practical method that combines the strengths of RL and MPC for ASU control in DR scenarios. Specifically, we investigate a hierarchical framework that integrates the scheduling decisions of an RL agent with a lower-level LMPC system. We opt for linear MPC over nonlinear MPC due to its faster inference time, ensuring it doesn't detract from RL's computational efficiency, though this comes at the cost of some control performance. This approach aims to leverage RL's learning capabilities and forward-inference efficiency while benefiting from MPC's stabilizing effect on the learning process.

\subsection{Related Works}
The DR problem for industrial processes falls within the broader class of enterprise-wide optimization (EWO) problems described by \citet{flores-tlacuahuac_simultaneous_2006}. In the context of DR, this often involves solving complex mixed-integer dynamic optimization (MIDO) problems, which combine discrete decisions (e.g., product scheduling) with continuous dynamic process models. These problems are typically transformed into mixed-integer nonlinear programming (MINLP) formulations for solution. While MIDO and MINLP approaches offer a rigorous framework for integrating scheduling and control decisions~\citep{zhang2015air}, they often face computational challenges due to problem size and complexity.

\citet{pattison_optimal_2016} introduced scale-bridging models (SBMs) for DR applications, using data-driven low-order dynamic models to ensure feasible schedules with reduced computational burden. The approach demonstrated significant cost savings when applied to an air separation unit under time-varying electricity prices. Building on this work, \citet{dias_simulation-based_2018}, \citet{caspari_integration_2020}, and \citet{schulze2023datadrivenmodelreductionnonlinear} compared different paradigms for integrating scheduling and control in DR problems. \citeauthor{dias_simulation-based_2018} developed a simulation-based optimization framework that combines SBMs with model predictive control (MPC), while \citeauthor{caspari_integration_2020} contrasted this ``top-down'' approach with a ``bottom-up'' economic MPC formulation. \citeauthor{schulze2023datadrivenmodelreductionnonlinear} developed a Koopman-based approach achieving real-time NMPC with 98\% reduced computational cost. Applied to an air separation unit, this method demonstrated ~8\% cost savings over steady-state operation, though highlighting tradeoffs between computational efficiency and economic optimality.

Reinforcement learning (RL) presents a compelling approach to demand response in industrial processes, offering advantages over traditional optimization methods. RL learns optimal control policies through environment interaction without requiring explicit mathematical models of system dynamics or constraints. Its effectiveness has been demonstrated in industrial applications, particularly in fed-batch bioreactors \citep{kaisare2003simulation, peroni2005optimal}, which present significant run-to-run variabilities \citep{YOO2021108, PETSAGKOURAKIS202235}. Recent advances include integrating RL with PID controllers for improved interpretability and sample efficiency \citep{lawrence2022deep}, with \cite{CIRL} showing enhanced performance through combined neural network-PID architectures. The data-driven nature of RL complements scheduling-based modeling approaches by leveraging historical data to improve decision-making. While challenges such as sample efficiency and stability remain, modern approaches such as scalable algorithms through action-space dimensionality reduction \citep{zhu2020scalable} and hierarchical RL frameworks have shown promise in industrial control applications, as demonstrated by \cite{kim_optimal_2023} in optimal continuous control of refrigeration systems under Time-of-Use pricing. While these previous works have demonstrated the potential of both traditional optimization and reinforcement learning approaches, the integration of RL with MPC for demand response optimization remains largely unexplored. Our work addresses this gap by presenting a hierarchical framework that combines reinforcement learning with LMPC, demonstrating enhanced sample efficiency and constraint satisfaction while maintaining economic performance in industrial demand response applications.

\section{Methodology}
\subsection{Reinforcement Learning for Control}\label{sec:rl-control}

In the RL framework, an agent learns to make decisions by interacting with an environment, aiming to maximize a cumulative reward signal.
Formally, RL problems are often modeled as Markov Decision Processes (MDPs), defined by the tuple $\langle \mathcal{X}, \mathcal{U}, f, r, \gamma \rangle$. Here, $\mathcal{X}$ represents the state space, and $\mathcal{U}$ is the control input space. The transition function $f: \mathcal{X} \times \mathcal{U} \times \mathcal{X} \rightarrow [0,1]$ describes the (stochastic) dynamics of the environment, while the reward function $r: \mathcal{X} \times \mathcal{U} \times \mathcal{X} \rightarrow \mathbb{R}$ quantifies the desirability of transitions. The discount factor $\gamma \in [0,1]$ balances immediate and future rewards.

The goal in RL is to find the policy $\pi^*: \mathcal{X} \rightarrow p(\mathcal{U})$ that maximizes the expected cumulative discounted reward:
\begin{equation}
    \pi^* = \max_{\pi} \mathbb{E}_{\tau \sim\pi}\left[\sum_{t=0}^{T} \gamma^t r_t(\mathbf{x}_t, \mathbf{u}_t, \mathbf{x}_{t+1}) \right]
\end{equation}
where $\mathbf{x}_t, \mathbf{x}_{t+1} \in \mathcal{X}$, $\mathbf{u}_t \in \mathcal{U}$, and the expectation is taken over the trajectories induced by the policy $\pi$ and the environment dynamics $f$.

\subsection{RL Control Structure}\label{sec: RL_Constrol_Struct}
In this work, we explore two methodologies for applying RL to process control: a direct formulation and a control-informed approach. The following are formulated in a deterministic setting, in terms of both the dynamics, $\mathbf{x}_{t+1} = F(\mathbf{x}_t, \mathbf{u}_t)$, and the control policy, $\mathbf{u}_t = \pi(\mathbf{x}_t)$ .

\begin{figure}[h]
    \centering
    \begin{tikzpicture}[auto, node distance=0.7cm,>=latex']
        \tikzstyle{block} = [draw, fill=blue!20, rectangle, 
            minimum height=2em, minimum width=4em]
        
        \node [coordinate] (input1) {};
        \node [block, below=of input1] (rl1) {RL Agent};
        \node [block, below=of rl1] (lmpc) {LMPC};
        \node [block, below=of lmpc] (system1) {System};
        \node [coordinate, below=of system1] (output1) {};
        
        \draw [->] (rl1) -- node[right] {$\mathbf{x}^*_t$} (lmpc);
        \draw [->] (lmpc) -- node[right] {$\mathbf{u}_t$} (system1);
        \draw [->] (system1) -- ++(-1cm,0) |- node[left] {$\mathbf{x}_{t+1}$} (lmpc);
        \draw [->] (lmpc) -- ++(-1cm,0) |- node[left] {} (rl1);
        
        \node [coordinate, right=3cm of input1] (input2) {};
        \node [block, below=of input2] (rl2) {RL Agent};
        \node [block, below=2.15cm of rl2] (system2) {System};
        \node [coordinate, below=of system2] (output2) {};
        
        \draw [->] (rl2) -- node[right] {$\mathbf{u}_t$} (system2);
        \draw [->] (system2) -- ++(-1cm,0) |- node[left] {} (rl2);
     
        \node[left] at ($(system2)+(-1cm,1.5cm)$) {$\mathbf{x}_{t+1}$};
        
        \node [below=-0.3cm of output1] {(a) With LMPC};
        \node [below=-0.3cm of output2] {(b) Without LMPC};
    \end{tikzpicture}
    \vspace{-2mm}
    \caption{RL-based control system diagrams: (a) with LMPC and (b) without LMPC}
    \label{fig:control-systems}
\end{figure}
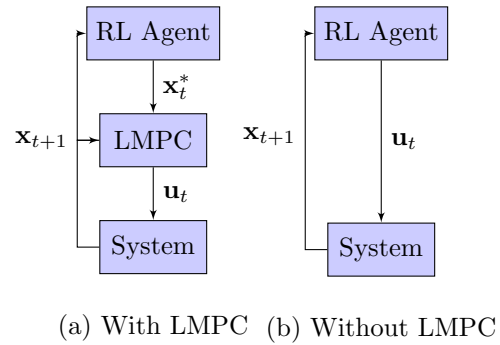
In the direct, control-agnostic formulation, we train an RL agent to directly output control actions. The optimization problem can be formulated as:
\begin{align*}
    \max_{\pi} \quad &\sum^T_{t = 0}  \gamma^t r_t(\mathbf{x}_t, \mathbf{u}_t = \pi(\mathbf{x}_t), \mathbf{x}_{t+1}) \\
    \text{s.t.} \quad &  \mathbf{x}_0 = \mathbf{x}(t_0) \\
    &\mathbf{x}_{t+1} = F(\mathbf{x}_t, \mathbf{u}_t = \pi(\mathbf{x}_t)) \\
\end{align*}
where $\mathbf{x}_t$ are the differential states, and the deterministic system dynamics $F: \mathcal{X} \rightarrow \mathbb{R}^{n_{x}}$ correspond to the process model, which is represented as a semi-explicit differential algebraic equation (DAE) system. 

The hierarchical, control-informed approach combines the strengths of RL and LMPC. This formulation assumes the availability of a linear model of the plant, typically obtained from previous control experiments or system identification. In this framework, the RL agent learns to provide setpoints to a lower-level LMPC system. The optimization problem becomes:
\begin{align*}
    \max_{\pi} \quad &\sum^T_{t=0}\gamma^t r_t(\mathbf{x}_t, \mathbf{u}_t, \mathbf{x}_{t+1}) \\
    \text{s.t.} \quad &\mathbf{x}_0 = \mathbf{x}(t_0) \\
    &\mathbf{u}_t = \psi(\mathbf{x}_{t-1}, \mathbf{x}^*_{t-1} = \pi(\mathbf{x}_{t-1})) \\
    &\mathbf{x}_{t+1} = F(\mathbf{x}_t, \mathbf{u}_t) \\
\end{align*}
In this case, $\pi$ represents the RL policy that provides setpoints $\mathbf{x}^*$ to the control law $\psi$ (written explicitly for simplicity), which then in turn determines the control actions $\mathbf{u}_t$. In this work, the control law $\psi$ is represented by a lower-level LMPC optimization that solves a quadratic program at each timestep to track the RL-provided setpoints. We employ the process model, linear dynamics, and cost matrices provided by \citet{dias_simulation-based_2018}. While LMPC is chosen for its computational efficiency, its linear approximation of the nonlinear ASU dynamics can limit control performance, particularly during large transitions. 
\subsection{Constraint Handling}
While recent works investigate directly embedding constraints in RL~\citep{burtea2024constrained}, here we simply formulate operational constraints of the problem as penalties to the reward function. For a path constraint of the form $g_i(\mathbf{x}) \leq 0$, the penalty at time $t$ is defined as:
\begin{equation} \label{eq:pathcon1}
r_{t, path} = \begin{cases}
\lambda g_i(\mathbf{x_t}) & \text{if } g_i(\mathbf{x}) > 0 \\
0 & \text{if } g_i(\mathbf{x}) \leq 0
\end{cases}
\end{equation}
where $g_i(\mathbf{x})$ represents the nonlinear constraint function and $\lambda$ represents a scaling factor to weight the constraints. For a terminal constraint of the form $h_i(\mathbf{x}_T) \leq 0$, we can densify the reward signal by relaxing it to a path constraint that activates when $t > t_a$:
\begin{equation} \label{eq:pathcon2}
r_{t, path} = \begin{cases}
\lambda (h_i(\mathbf{x}_t))^2 & \text{if }h_i(\mathbf{x}_t) > 0 \text{ and } t > t_a \\
0 & \text{otherwise}
\end{cases}
\end{equation}
where $h_i(\mathbf{x}_t)$ represents the terminal constraint function evaluated at time $t$, and $t_a$ represents the activation time, selected to balance  effective load shifting influence and a sufficient response time for the agent. This modification provides continuous feedback throughout the latter portion of the episode instead of only at the terminal time $T$. Equality constraints can be treated by duplicating \eqref{eq:pathcon1}--\eqref{eq:pathcon2}.

\subsection{Policy Optimization}
Policy optimization is a fundamental approach in RL for finding an optimal policy. In continuous action spaces, policy gradient methods have shown significant promise. Among these, the Deep Deterministic Policy Gradient (DDPG) algorithm has emerged as particularly effective for continuous control tasks~\citep{lillicrap2019continuouscontroldeepreinforcement}, and we employ this algorithm for policy optimization throughout this work, although the proposed method is agnostic to the choice of RL algorithm.
DDPG combines elements of both policy-based and value-based methods, utilizing an actor-critic architecture. Central to this approach is the concept of the Q-function, also known as the action-value function. The Q-function is defined as the expected cumulative discounted reward when taking action $\mathbf{u}$ in state $\mathbf{x}$ and following policy $\pi$ thereafter:
\begin{equation}
Q^\pi(\mathbf{x}_t, \mathbf{u}_t) = \mathbb{E}_\pi\left[\sum_{t=0}^{T} r_t | \mathbf{x}_0 = \mathbf{x}, \mathbf{u}_0 = \mathbf{u}\right]
\end{equation}
where $r_t$ is the reward at time step $t$.
In complex environments with continuous or high-dimensional state and action spaces, the Q-function can be approximated using a neural network:
\begin{equation}
Q(\mathbf{x}, \mathbf{u}) \approx Q_\theta(\mathbf{x}, \mathbf{u})
\end{equation}
where $\theta$ represents the parameters of the neural network.
Building on these concepts, DDPG learns a deterministic policy $\mu: \mathcal{X} \rightarrow \mathcal{U}$ and an action-value function $Q: \mathcal{X} \times \mathcal{U} \rightarrow \mathbb{R}$ simultaneously. The objective is to maximize the expected return $J(\mu_{\theta^\mu})$ defined as:
\begin{equation}
J(\mu_{\theta^\mu}) = \mathbb{E}_{\mu_{\theta^\mu}}\left[\sum^T_{t=0}r_t(\mathbf{x_t}, \mathbf{u_t}, \mathbf{x}_{t+1})\right]
\end{equation}
where $\mu$ is the deterministic policy. The policy is updated using the deterministic policy gradient theorem:
\begin{multline}
\nabla_{\theta^\mu} J(\mu_{\theta^\mu}) \approx \mathbb{E}_{\textbf{x} \sim \rho^\beta} [\nabla_u Q(\textbf{x},\textbf{u}|\theta^Q)|u= \\ \mu(\textbf{x}|\theta^\mu)  \nabla_{\theta^\mu} \mu(\textbf{x}|\theta^\mu)]
\end{multline}
where $\theta^\mu$ are the parameters of the policy network, $\theta^Q$ are the parameters of the Q-function network, and $\rho^\beta$ is the state distribution under the behavior policy $\beta$.
\vspace{-3mm}
\section{Computational Case Study}
\begin{figure}[t]
    \centering
     \includegraphics[clip, trim=6.9cm 9.1cm 10cm 4.3cm, width=0.5\textwidth]{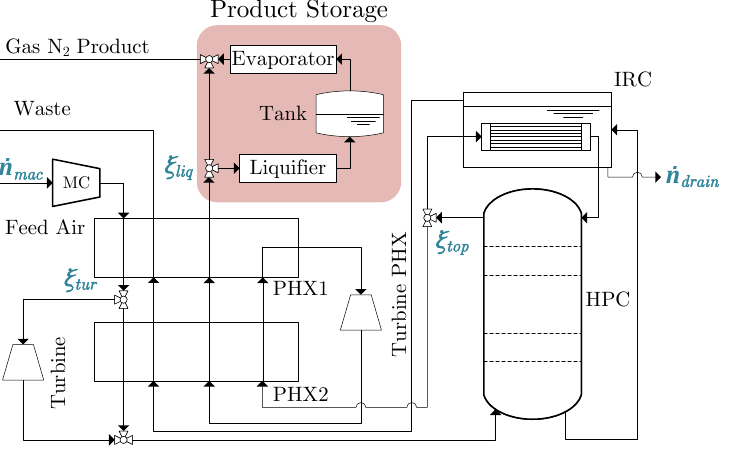}
    \vspace{-3mm}
    \caption{ASU Process Flowsheet. Manipulated Variables are marked in blue and the Product Storage Section is shaded in Red}
    \label{fig:ASU_Process_Flowsheet}
\end{figure}

The ASU used as the case study within this work is an openly available benchmark process for flexible operation problems \citep{tsay2020benchmark} and is depicted in Figure \ref{fig:ASU_Process_Flowsheet}. This benchmark process represents the environment that both agents can interact with. The ASU's main process unit is the single cryogenic distillation column which produces high-purity Nitrogen product. The feed air ($\dot{n}_{mac}$) is passed through an electric compressor (MC in Figure~\ref{fig:ASU_Process_Flowsheet}) and is cooled in the first Primary Heat Exchanger (PHX1) after which a fraction $\xi_{tur}$ is sent to the turbine to generate electricity. The rest of the feed is sent to the second Primary Heat Exchanger 2 (PHX2). This liquified stream is combined with the turbine outlet to feed the High Pressure Column (HPC). The bottoms stream is then sent to the reboiler side of the Integrated Reboiler/Condenser (IRC) Unit. The high-purity nitrogen from the distillate is split by fraction $\xi_{top}$ and sent to the primary heat exchanger. To enable the flexible operation of the ASU, nitrogen product can be liquified and sent to the storage tank. This stored product can be evaporated to meet the instantaneous demand when the production rate is decreased allowing the shifting of load. The fraction sent to the liquified $\xi_{liq}$ is set with the following control logic so that the product demand is met at all times:
\begin{align}
\xi_{\text{liq}} = \begin{cases}
1 - \frac{\dot{n}_{\text{demand}}}{\dot{n}_{\text{product}}} & \text{when } \dot{n}_{\text{product}} > \dot{n}_{\text{demand}} \\
0 & \text{otherwise}
\end{cases} 
\end{align}
For a detailed explanation of the modeling of the ASU, we direct the reader to the work by \citet{caspari_integration_2020}.
\subsection{Operational Constraints}
The ASU system must be controlled to satisfy certain operational constraints regarding product purity, heat exchanger minimum temperature difference, and vessel holdup. Along with these constraints is a terminal constraints on the storage tank holdup. The storage constraint in particular is enforced to ensure enough storage for the following days. We note that this constraint is artificial since, over a long time horizon, it may be optimal to finish the horizon at lower or higher storage holdups dependent on the following day's electricity profile and demand. The specific enforced path and terminal constraints are shown in Table \ref{tab:operational_cons} and the manipulated variable bounds are shown in Table \ref{tab:MV_bounds}. 

\begin{table}[h]
\begin{center}
\caption{ASU Operational Constraints}
\label{tab:operational_cons}
\scalebox{0.9}{
\begin{tabular}{@{}ccc@{}}
\toprule
Variable                  & Constraint                           & Type     \\ \midrule
$I_{product}$ {[}ppm{]}    & $0 \leq I_{product} \leq 1500$         & Path     \\
$\Delta T_{IRC}$ {[}K{]} & $2 \leq \Delta T_{IRC} \leq 5$       & Path     \\
$N_{tank}$ {[}mol{]}     & $ 864000 \leq N_{tank} \leq 3456000$ & Path     \\
$F_{tank}$ {[}mol/s{]}     & $0 \leq F_{tank}$                    & Path     \\
$N_{tank}$ {[}mol          & $N_{tank} = 1728000$                  & Terminal \\ \bottomrule
\end{tabular}}
\end{center}
\end{table}

\begin{table}[h]
\begin{center} 
\caption{Manipulated Variable Bounds}
\label{tab:MV_bounds}
\begin{tabular}{@{}cc@{}}
\toprule
MV                          & Bounds                            \\ \midrule
$\dot{n}_{MAC} ${[}mol/s{]} & $ 30 \leq \dot{n}_{MAC} \leq 50 $ \\
$\xi_{tur}$ {[}-{]}         & $0 \leq \xi_{tur} \leq 0.1$       \\
$\xi_{liq}$ {[}-{]}         & $0 \leq \xi_{liq} \leq 1$         \\
$\xi_{top}$ {[}-{]}         & $0.51 \leq \xi_{top} \leq 0.54$   \\
$F_{drain}$ {[}mol/s{]}       & $0 \leq F_{drain} \leq 2$           \\ \bottomrule
\end{tabular}
\end{center}
\end{table}
\vspace{-3mm}
\subsection{Reward Function}
\vspace{-3mm}
The objective of the case study is to minimize operational costs while also satisfying operational constraints. To translate this to the RL framework, we must construct a reward function that describes this goal. The economic half of the objective can be described as the maximization of the negative electricity costs as follows: 
\begin{equation}
r_{t,elec} = - \left( p_{t, elec}(P_{comp}+P_{liq}-P_{tur})\Delta t \right) 
\end{equation}
where $p_{t, elec}$ [\$/MWh] the electricity price at time $t$, and $P_i$ [MW] is the power consumed/generated by unit $i$ at time step $t$.  The total reward at each timestep can then be written as:
\begin{equation}
r_t = r_{t,elec} + \sum^{n_g}_i r_{t,path,i} + r_{t,terminal}
\end{equation}
\vspace{-3mm}
\subsection{State and Action Spaces}
The RL agent's state space $\mathcal{X}$ is designed to contain the relevant information to make decisions. For both RL agents described in Section \ref{sec: RL_Constrol_Struct} the state is defined as:
\begin{equation}
    \mathbf{x}_t = \{I_{product}, \Delta T_{IRC}, N_{tank}, F_{tank}, p_t,..., p_{t+11}, t_d\} \in \mathbb{R}^{17}
\end{equation}
where $I_{product}$ [ppm] is the product impurity, $\Delta T_{IRC}$ [K] is the temperature difference in the internal rectification column, $N_{tank}$ [mol] is the amount of product in the storage tank, $F_{tank}$ [mol/s] is the flow rate to the storage tank, $p_t$ [\$/MWh] represents the electricity prices for the next 12 hours (perfect forecast is assumed), and $t_d \in [0,24]$ represents the time within the day.

\begin{figure}[b]
    \centering
    \includegraphics[clip, trim=0.1cm 0.4cm 0cm 0.7cm, width=0.8\columnwidth]{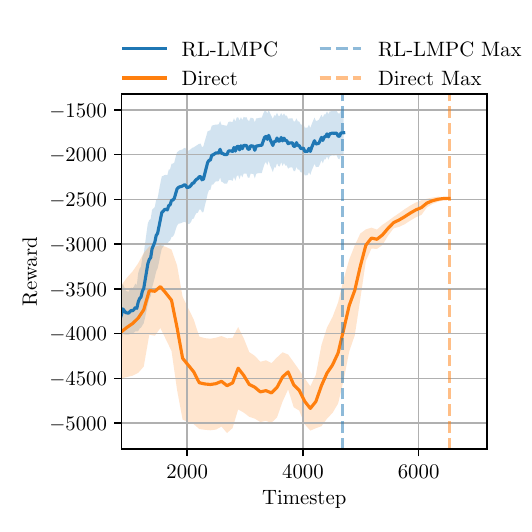}
    \vspace{-3mm}
    \caption{Learning curve for both the RL-LMPC and Direct agents with rolling mean and variance, truncated when the maximum reward is reached within the 10,000 timestep budget}
    
    \label{fig:learning-curve}
\end{figure}

For the direct RL agent, the action space consists of all manipulated variables:
\begin{equation}
\mathbf{u}_t^{\text{direct}} = \{\dot{n}_{MAC}, \xi_{tur}, \xi_{top}, F_{drain}\} \in \mathbb{R}^4
\end{equation}
where $\dot{n}_{MAC}$ [mol/s] is the main air compressor flow rate, $\xi_{tur}$ [-] is the turbine split fraction, $\xi_{top}$ [-] is the top column split fraction, and $F_{drain}$ [mol/s] is the reboiler drain flow rate.
For the RL-LMPC agent, the action space is reduced to a single setpoint:
\begin{equation}
\mathbf{u}_t^{\text{LMPC}} = \{\dot{n}_{product,sp}\} \in \mathbb{R}
\end{equation}
where $\dot{n}_{product,sp}$ [mol/s] is the setpoint for the product flow rate. Since the LMPC can effectively maintain the other three controlled variables around their steady-state values through coordinated manipulation of the lower-level inputs, we can reduce the action space from $\mathbb{R}^4$ to $\mathbb{R}$. This simplification of the action space dimensionality allows the RL agent to focus solely on the critical production rate decisions while relying on the LMPC for stable operation of the remaining process variables.
\section{Results and Discussion}

\begin{figure}[b!]
    \centering
    \includegraphics[clip, trim=0.1cm 0.4cm 0cm 0.52cm,width=0.75\linewidth]{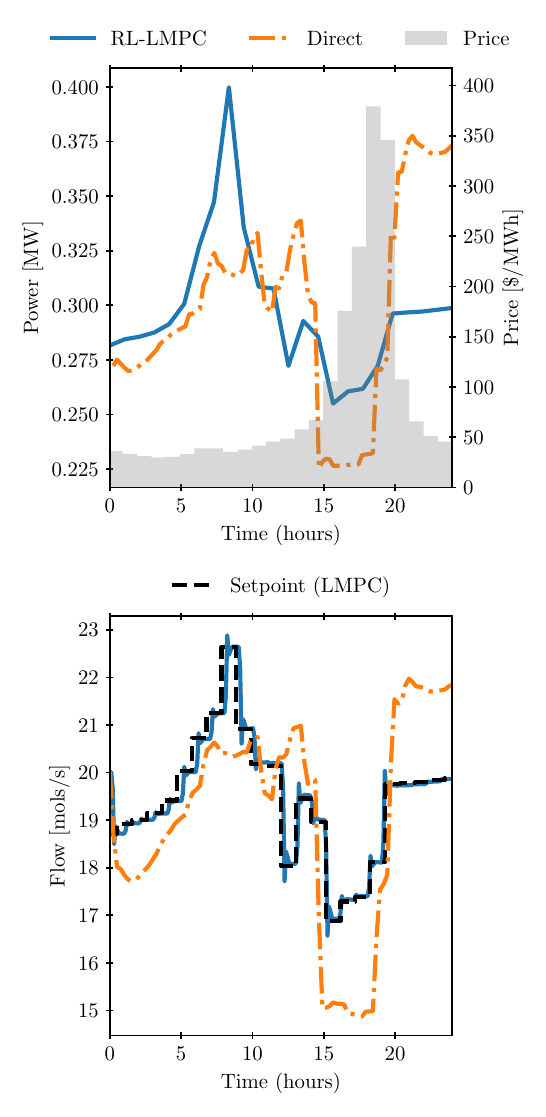}
    \vspace{-3mm}
    \caption{Top: power demand and price profiles. Bottom: production rate for the RL-LMPC (blue, solid lines) and direct (Orange, dash-dot lines) agents.}
    \label{fig:power-flow}
\end{figure}

\begin{figure*}[t!]
    \centering
    \includegraphics[clip, trim=0.1cm 0.4cm 0cm 0.52cm, width = 0.8\linewidth]{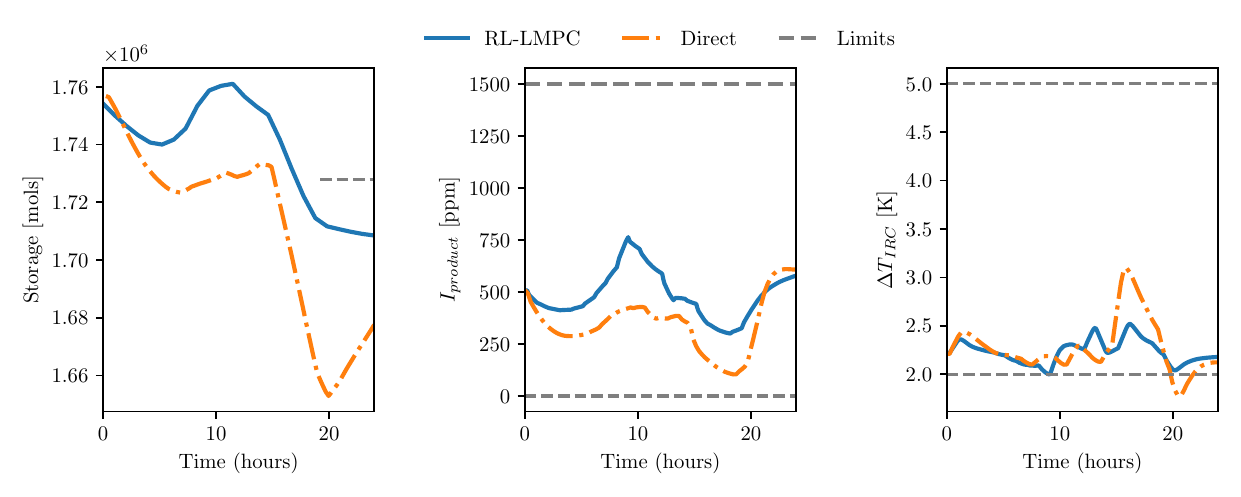}
    \vspace{-3mm}
    \caption{Storage, Product Impurity, and IRC Temperature Difference or the RL-LMPC (blue, solid lines) and direct (Orange, dash-dotted lines) agents with the imposed constraints (gray, dashed lines)}
    \label{fig:storage}
\end{figure*}

Figure \ref{fig:learning-curve} presents the learning curves for both RL-control approaches. Both algorithms were implemented using the Stable-Baselines3 library by \cite{stable-baselines3}, with default hyperparameters and trained with a budget of 10,000 timesteps and episodes of 96 timesteps, corresponding to a single day.
The RL-LMPC framework demonstrates more efficient learning by converging to a high-quality policy after approximately 4000 timesteps, while the direct RL method required nearly 7000 timesteps to improve its final performance. This faster learning rate of the RL-LMPC can be attributed to the embedded LMPC reducing the exploration space by handling constraints explicitly, thereby allowing the RL agent to focus on optimizing the control policy within a feasible operating region. Moreover, the dimensionality of the explored action space is reduced considerably. After training, the best policy achieved within the 10,000 timestep budget was fixed and used to generate the subsequent operational trajectories shown in Figures \ref{fig:power-flow} and \ref{fig:storage}.

Figure~\ref{fig:power-flow} first shows the power demand of the ASU unit overlayed over the electricity price profile and the product flow rates for both direct and RL-LMPC agents.  Both control policies exhibit load-shifting behavior, strategically reducing power consumption during periods of peak electricity prices and increasing production during lower-price periods. The direct RL policy adopts a more conservative approach during the first 10 hours of operation, maintaining power consumption between 0.300-0.330 MW, despite relatively favorable electricity prices. In contrast, within the same period, the RL-LMPC increases production in preparation for the high energy prices. During the latter half of the operating period (hours 10-20), the direct policy becomes more aggressive, showing significant  variations in power consumption in response to price fluctuations and the terminal storage constraint. While the RL-LMPC demonstrates effective load-shifting, some of its responses appear more aggressive than necessary, which may be attributed to the current LMPC tuning parameters rather than fundamental limitations of the approach. This follows the observations of \citet{caspari_integration_2020} that allowing a scheduling layer to control the setpoints directly can give the flexibility to enable more aggressive behavior. 

Figure \ref{fig:storage} presents the trajectories of three constrained variables: storage levels, product impurity, and IRC temperature differences, comparing the performance of RL-LMPC and direct RL control approaches. The storage trajectory is subject to a terminal constraint requiring a return to the mid-level by the end of the 24-hour period. Both control approaches fail to satisfy this terminal condition, though with markedly different magnitudes. The direct RL agent exhibits substantially larger deviations, primarily due to its conservative operating strategy during initial hours, which fails to build a sufficient storage buffer for subsequent operations. The RL-LMPC agent, while still violating the terminal constraint, maintains closer proximity to the desired final state.

In terms of path constraints, both agents successfully maintain product impurity within the specified bounds throughout the operational period. However, a significant performance divergence occurs in the IRC temperature difference at hour 20, where the direct RL agent notably violates the lower bound. The RL-LMPC agent, benefiting from its embedded linear constraint model in the lower-level LMPC, successfully avoids such violations. This superior performance stems from the LMPC's explicit constraint handling mechanism, which reduces constraint-violating actions from the RL agent's exploration space, demonstrating a key advantage of the hybrid approach in managing operational constraints.


\section{Conclusions and Future Works}
This work investigates integrating reinforcement learning with model predictive control for demand response scheduling in air separation units. The proposed RL-LMPC framework showed certain advantages over a direct RL implementation in terms of sample efficiency and constraint satisfaction. By incorporating the LMPC's knowledge of system dynamics and constraints, the hybrid approach reduces the solution space for the RL agent, helping avoid some infeasible operating conditions. This reduction in searchable space contributed to more efficient learning and we found the hybrid approach to exhibit more stable control behavior. Both approaches implemented load-shifting strategies in response to electricity price variations, with the RL-LMPC framework showing better capability at maintaining feasible operation.

Future work may investigate fully RL-based hierarchical control structures, e.g., where both high-level and low-level controllers are RL agents. The framework can be extended to incorporate stochastic electricity price distributions and varying initial storage levels. These developments may help advance the practical implementation of RL-based strategies in industrial process scheduling.

\bibliography{ifacconf}             

\end{document}